\documentclass[10pt]{article}

\usepackage{blindtext}
\usepackage{graphicx}
\usepackage[fleqn]{amsmath}
\usepackage{amsmath}
\usepackage[T1]{fontenc}
\usepackage[utf8x]{inputenc}
\usepackage{bbold}
\usepackage{epstopdf}
\usepackage{breqn}
\usepackage{mathtools}
\newcommand{\sign}{\text{sign}}
\usepackage{scicite}
\usepackage{caption}
\usepackage{titlesec}

\newcommand{\diagentry}[1]{\mathmakebox[1.8em]{#1}}
\newcommand{\xddots}{%
  \raise 4pt \hbox {.}
  \mkern 6mu
  \raise 1pt \hbox {.}
  \mkern 6mu
  \raise -2pt \hbox {.}
}
\usepackage{subfig}
\usepackage{setspace}
\AtBeginDocument{%
  \addtolength\abovedisplayskip{-0.5\baselineskip}%
  \addtolength\belowdisplayskip{-0.5\baselineskip}%
}

\usepackage{times}

\newenvironment{sciabstract}{%
\begin{quote} \bf}
{\end{quote}}

\newcounter{lastnote}

\title{Study of the Sparse Superposition Codes and the Generalized 
Approximate Message Passing Decoder for the Communication over
Binary Symmetric and Z Channels}

\author
{Alper Köse, Berke Aral Sönmez\\
\\
\normalsize{Department of Electrical Engineering, Ecole Polytechnique Federale de Lausanne}\\
}

\date{}

\begin{document} 


\baselineskip24pt


\maketitle

\begin{sciabstract}
  In this project, the behavior of Generalized Approximate Message-Passing Decoder for BSC and Z Channel is studied using i.i.d matrices for constructing the codewords. The performance of GAMP in AWGN Channel is already evaluated in the previous scientific work of Jean Barbier[1], therefore, this project mainly focuses on the performance of GAMP decoder for BSC and Z Channel. We evaluate the performance of the GAMP decoder for sparse superposition codes at various settings and compare the performance of decoder for different channels and parameters.
\end{sciabstract}

\section{Introduction}

\subsection{Sparse Superposition Codes}

We can define sparse superposition code as a signal that consists of multiple sections and each section is a vector with length \(B\) where \(B-1\) of the values in a section are \(0\)s. For this project, we used \(B\) values where \(B\) is an integer power of 2. For instance, in case of \(B=4\), the alphabet has symbols \(1000, 0100, 0010, 0001\) and it can be said that the number of possible letters in the alphabet equals to \(B\). If we want to send a message \(\mathbf{x}\) with \(L\) sections, the total length of the message becomes \(N=BL\) and the sparsity of this vector increases when \(B\) increases.  To send the sparse superposition codes through the channel, we multiply the message \(\mathbf{x}\) by a matrix \(\mathbf{A}\) and obtain a vector \(\mathbf{z}\) which is \(\mathbf{z}=\mathbf{A}\mathbf{x}\) and send \(\mathbf{z}\) through the noisy channel. \(\mathbf{A}\)  is a transform matrix with size \(MN\)  and the length of the obtained \(\mathbf{z}\) vector is equal to \(M\). For this project, we used a random gaussian matrix with 0 mean and variance \(\bigtriangleup\) as our \(\mathbf{A}\) matrix. However, we need to fix the power \(P\) of \(\mathbf{z}\) to one, therefore we created a Gaussian matrix with 0 mean and variance 1 and divided it by \(\sqrt{L}\) which becomes our \(\mathbf{A}\) matrix.
    
To determine the \(M\) values which allow this system to communicate through the channel we need to calculate our communication rate \(R\) with respect to \(M\). Number of informative bits in the message \(\mathbf{x}\) can be defined as \(K=L\log_2{B}\) and if we define a parameter called \(\alpha \) which is equal to \(\alpha=M/N\), the communication rate becomes \(R=K/M=L\log_2{B}/(N\alpha)=\log_2{B}/(B\alpha)\). Therefore, we can adjust the communication rate \(R\) by changing the number of rows in matrix \(\mathbf{A}\).

After sending the codeword \(\mathbf{z}\) through the AWGN channel we get an output \(\mathbf{y}\) which is \(\mathbf{y=z+N}\), where \(\mathbf{N}\) is the noise of the channel. However, BSC and Z channels are binary channels and only 0s and 1s can be sent through those channels and \(\mathbf{z}\) is not in a binary form. Therefore, instead of sending \(\mathbf{z}\), we send \(\sign \mathbf{(z)}\) information through the channel. In binary symmetric channel, input bits 0 and 1  have a flip probability of \(\epsilon\) whereas in Z channel, only 0s have a flip probability of \(\epsilon\) and 1s are always transmitted correctly.

Our main ambition after transmission is to decode the message \(\mathbf{x}\) correctly, with the help of the output \(\mathbf{y}\), transform matrix \(\mathbf{A}\) and the signal-to-noise ratio (SNR) or \(\epsilon\) of the channel. To calculate the error obtained after the transmission, we use two measures which are mean square error(MSE or \(\widetilde{E}\)) and section error rate (SER). We can compute MSE and SER as:
\begin{equation} 
	\widetilde{E} =\frac{1}{L}\sum_{i=1}^{N}(x_{i}-\hat{x_{i}})^2
\end{equation}
\begin{equation}
	SER=\frac{1}{L}\sum_{l=1}^{L}\mathbb{1}(x_{l}\ne\hat{x_{l}})
\end{equation}

By using SER and MSE values, it is possible to observe the behaviour of the decoder for different values of \(B\), \(R\), channel type and \(\epsilon\) or signal-to-noise ratio (SNR).

\subsection{Generalized Approximate Message Passing
Algorithm}

To decode the output vector \(\mathbf{y}\) to estimate the input \(\mathbf{x}\), we had to choose an algorithm to use. For this goal, after fruitful discussions with Jean Barbier in the beginning of the semester, we decided to use Generalized Approximate Message Passing (GAMP) algorithm instead of Belief Propagation (BP) since the factor graph associated to our estimation problem is densely connected (BP is used when the factor graph sparse and it is computationally inefficient to use in case of densely connected graph).
Generalized Approximate Message Passing  gives us an approximation, but it is known as an efficient method for estimating unknown vector in a linear mixing problem. Let us consider our case as an example: There is an unknown vector, \( \mathbf{x} \), which is a sparse superposition vector with the entries of every section are zero except one component. Firstly, this vector is multiplied with a transform matrix \( \mathbf{A} \), which is a random gaussian matrix in our case, therefore we get a codeword \( \mathbf{z} \), namely \(\mathbf{z}=\mathbf{A}\mathbf{x} \). After this operation, \( \mathbf{z} \) generates an output vector \( \mathbf{y} \) with respect to a conditional probability distribution  \( p_{Y |Z} (y |z ) \). The problem here is to estimate the input vector \( \mathbf{x} \) from the output vector \( \mathbf{y} \) and transform matrix \( \mathbf{A} \). Optimal estimation of \( \mathbf{x} \) is hard since the components of it are coupled in \( \mathbf{z} \). To achieve this goal, we use the GAMP algortihm, which is computationally simple, in Rangan's paper[2].

Given a matrix \( \mathbf{A}\in \mathbb{R}^{mxn} \), system inputs and outputs \( \mathbf{q} \) and \( \mathbf{y} \) and scalar estimation functions \(g_{in}(.)\) and \(g_{out}(.)\) generate a sequence of estimates  \(\hat{ \mathbf{x}} \)(t), \( \hat{\mathbf{z}} \)(t), for  \(t=0,1,...\) through the following recursions:

1) Initialization: Set \(t = 0\) and set \(\hat{x}_{j}(t) \) and \(\tau^{x}_{j}(t)\) to some initial values.

2) Output linear step: For each  \(i\), compute:
\begin{equation}
\tau^{p}_{i}(t)=\sum\limits_{i} |a_{ij}|^{2} \tau^{x}_{j}(t)
\end{equation}
\begin{equation}
\hat{p}_{i}(t)=\sum\limits_{j} a_{ij}\hat{x}_{j}(t)-\tau^{p}_{i}(t) \hat{s}_{i}(t-1)
\end{equation}
\begin{equation}
\hat{z}_{i}(t)=\sum\limits_{j} a_{ij}\hat{x}_{j}(t)
\end{equation}

where initially, we take  \( \hat{s}(-1) = 0. \)

3)Output nonlinear step: For each \( i \),
\begin{equation}
\hat{s}_{i}(t)=g_{out}(t,\hat{p}_{i}(t),y_{i}, \tau^{p}_{i}(t))
\end{equation}
\begin{equation}
	\tau^{s}_{i}(t)=-\frac{\partial}{\partial \hat{p}}g_{out}(t,\hat{p}_{i}(t),y_{i}, \tau^{p}_{i}(t))
\end{equation}

4)Input linear step: For each \( j \),
\begin{equation}
	\tau^{r}_{j}(t)=\left[\sum\limits_{i} |a_{ij}|^{2} \tau^{s}_{i}(t)\right]^{-1}
\end{equation}
\begin{equation}
	\hat{r}_{j}(t)=\hat{x}_{j}(t)+\tau^{r}_{j}(t)\sum\limits_{i} a_{ij}\hat{s}_{i}(t)
\end{equation}

5) Input nonlinear step: For each \( j \),
\begin{equation}
	\hat{x}_{j}(t+1)=g_{in}(t,\hat{r}_{j}(t),q_{j}, \tau^{r}_{j}(t))
\end{equation}
\begin{equation}
	\tau^{x}_{j}(t+1)=\tau^{r}_{j}(t)\frac{\partial}{\partial \hat{r}}g_{in}(t,\hat{r}_{j}(t),q_{j}, \tau^{r}_{j}(t))
\end{equation}

Then increment \( t=t+1 \) and return to step 2 until a sufficient number of iterations have been performed.

In our case, we do not use \( \mathbf{q} \), since it is used to derive the input \( \mathbf{x} \) in Rangan's paper, whereas we directly have input vector \( \mathbf{x} \) in the beginning.

To adapt GAMP algorithm to our problem, we had to derive \(g_{in}\) and \(g_{out}\) mathematically using the problem parameters. Then, our decoder would be ready for the simulations.

\section{Mathematical Calculations for GAMP}

\subsection{Calculation of \(g_{in}\)}

To find \(\mathbf{\hat{x}}\) in the fifth step of GAMP, we need to compute \(g_{in}(\hat{\mathbf{r}},q,\boldsymbol{\tau}^{r})\), equivalently \(\mathbb{E}(\mathbf{x}|\hat{\mathbf{r}},q,\boldsymbol{\tau}^{r})\).
\begin{equation}
\mathbb{E}(\mathbf{x}|\hat{\mathbf{r}},q,\boldsymbol{\tau}^{r})=\frac {\int\mathrm{d}\mathbf{x} p_{0}(\mathbf{x}) \mathcal{N}(\mathbf{x}|\mathbf{r},\boldsymbol{\tau})\mathbf{x}}      {\int\mathrm{d}\mathbf{x} p_{0}(\mathbf{x}) \mathcal{N}(\mathbf{x}|\mathbf{r},\boldsymbol{\tau}) }
\end{equation}

where 
\begin{equation}
p_{0}(\mathbf{x})=\frac{1}{B}\sum_{i=1}^{B} \delta_{x_{i},1} \prod_{k\neq i}^{B} \delta_{x_{k},0}
\quad \quad \quad  and \quad \quad \quad 
\mathcal{N}(\mathbf{x}|\mathbf{r},\boldsymbol{\tau})=  \frac{1}{\sqrt{(2\pi)^{B}|\boldsymbol{\Sigma}|}} e^{-\frac{1}{2} (\mathbf{x}-\mathbf{r})^{T}\boldsymbol{\Sigma}^{-1} (\mathbf{x}-\mathbf{r})}
\end{equation}

where 
\begin{equation}
\Sigma=
\begin{pmatrix}
\diagentry{{\tau}_{1}}\\
&\diagentry{{\tau}_{2}}\\
&&\diagentry{\xddots}\\
&&&\diagentry{{\tau}_{B}}\\
\end{pmatrix}
\end{equation}

In the equations, \(\mathbf{r}\) and \(\boldsymbol{\tau}\) are B-dimensional vectors in which \(\mathbf{r}\) gives the mean and \(\boldsymbol{\tau}\) values constitute the covariance matrix of the gaussian pdf.

The results of the \(\mathbb{E}(\mathbf{x}|\hat{\mathbf{r}},q,\boldsymbol{\tau}^{r})\) are computed analytically in Matlab using the information above. After this computation, we need to compute \(\boldsymbol\tau^{x}\). For this, we must find 
\(\boldsymbol\tau^{r}\frac{\partial}{\partial \boldsymbol{\hat{r}}}g_{in}(\boldsymbol{\hat{r}},q, \boldsymbol\tau^{r})\), equivalently \(var(\mathbf{x}|\hat{\mathbf{r}},q,\boldsymbol{\tau}^{r})\).
\begin{equation}
var(\mathbf{x}|\hat{\mathbf{r}},q,\boldsymbol{\tau}^{r})=\mathbb{E}(\mathbf{x}^{2}|\hat{\mathbf{r}},q,\boldsymbol{\tau}^{r})-\mathbb{E}(\mathbf{x}|\hat{\mathbf{r}},q,\boldsymbol{\tau}^{r})^{2}
\end{equation}

since, \(\mathbf{x}\) consists of 0s and 1s,  \(\mathbb{E}(\mathbf{x}^{2}|\hat{\mathbf{r}},q,\boldsymbol{\tau}^{r})=\mathbb{E}(\mathbf{x}|\hat{\mathbf{r}},q,\boldsymbol{\tau}^{r})\). Therefore, we can conclude that:
\begin{equation}
\boldsymbol\tau^{x}=\mathbf{\hat{x}}-\mathbf{\hat{x}}^{2}
\end{equation}

where the multiplication \(\mathbf{\hat{x}}^{2}=\mathbf{\hat{x}}\mathbf{\hat{x}}\) is element-wise.

\subsection {Calculation of \(g_{out}\)}
\subsubsection{\(\mathbf{\hat{s}}\) and \(\boldsymbol{\tau}^s\) for AWGN Channel}
    For the AWGN channel, we defined \(\mathbf{\hat{s}}\) as:
\begin{equation}
\mathbf{\hat{s}}=g_{out}(\mathbf{\hat{p}},\mathbf{y},\boldsymbol{\tau}^p)=\frac{(\mathbf{y}-\mathbf{\hat{p}})}{(\boldsymbol{\tau}^p+\tau^w)}
\end{equation}
where we define \(\tau^w\) as \(1/snr\). In the GAMP algorithm we defined \(\boldsymbol{\tau}^s\) as
\begin{equation}
\boldsymbol{\tau}^s=-\frac{\partial}{\partial \mathbf{\hat p}}(g_{out}(\mathbf{\hat{p}},\mathbf{y},\boldsymbol{\tau}^p))=-\frac{\partial}{\partial \mathbf{\hat p}}\frac{(\mathbf{y}-\mathbf{\hat{p}})}{(\boldsymbol{\tau}^p+\tau^w)}=\frac{1}{(\boldsymbol{\tau}^p+\tau^w)}
\end{equation}
\subsubsection{Computing \(\mathbf{\hat{s}}\) and \(\boldsymbol{\tau}^s\) for BSC}

While discussing GAMP algorithm, we said that \(\mathbf{\hat{s}}(t)=g_{out}(t,\mathbf{\hat{p}}(t),\mathbf{y}, \boldsymbol{\tau}^{p}(t))\) and for sum-product GAMP, we define \(g_{out}(\mathbf{\hat{p}},\mathbf{y},\boldsymbol{\tau}^p)\) as:
\begin{equation}
g_{out}(\mathbf{\hat{p}},\mathbf{y},\boldsymbol{\tau}^{p})=(\mathbf{\hat{z}}^0-\mathbf{\hat{p}})/\boldsymbol{\tau}^p
\end{equation}
where \(\mathbf{\hat{z}}^0:=\mathbb{E}(\mathbf{z}|\mathbf{\hat{p}},\mathbf{y},\boldsymbol{\tau}^p)\), \(Y\sim P_{Y|Z}\) and \(Z\sim \mathcal{N}(\mathbf{\hat{p}},\boldsymbol{\tau}^p)\). For BSC, we can define \(P_{Y|Z}\) as:
\begin{equation}
P_{Y|Z}=(1-\epsilon)\delta(\mathbf{y}-\pi(\mathbf{z}))+\epsilon\delta(\mathbf{y}+\pi(\mathbf{z}))
\end{equation}
where \(\epsilon\) is the flip probability and \(\pi(\mathbf{z})=sign(\mathbf{z})\). We can compute \(\mathbf{\hat{z}}^0\) such that:
\begin{equation}
\mathbf{\hat{z}}^0=\frac{1}{\mathcal{Z}}\int_{-\infty}^{\infty}\frac{\mathbf{z}e^{-(\mathbf{z}-\mathbf{\hat{p}})^2/2\boldsymbol{\tau}^p}}{\sqrt{2\pi \boldsymbol{\tau}^p}}((1-\epsilon)\delta(\mathbf{y}-\pi(\mathbf{z}))+\epsilon\delta(\mathbf{y}+\pi(\mathbf{z})))d\mathbf{z}
\end{equation}

\begin{multline}
 \mathbf{\hat{z}}^0=\frac{1}{\mathcal{Z}}\int_{-\infty}^{0}\frac{\mathbf{z}e^{-(\mathbf{z}-\mathbf{\hat{p}})^2/2\boldsymbol{\tau}^p}}{\sqrt{2\pi \boldsymbol{\tau}^p}}((1-\epsilon)\delta(\mathbf{y}+1)+\epsilon\delta(\mathbf{y}-1))d\mathbf{z}+\\ \frac{1}{\mathcal{Z}}\int_{0}^{\infty}\frac{\mathbf{z}e^{-(\mathbf{z}-\mathbf{\hat{p}})^2/2\boldsymbol{\tau}^p}}{\sqrt{2\pi \boldsymbol{\tau}^p}}((1-\epsilon)\delta(\mathbf{y}-1)+\epsilon\delta(\mathbf{y}+1))d\mathbf{z}
\end{multline}
We can write \(\mathcal{Z}\) as:
\begin{multline}
\mathcal{Z}=\int_{-\infty}^{0}\frac{e^{-(\mathbf{z}-\mathbf{\hat{p}})^2/2\boldsymbol{\tau}^p}}{\sqrt{2\pi \boldsymbol{\tau}^p}}((1-\epsilon)\delta(\mathbf{y}+1)+\epsilon\delta(\mathbf{y}-1))d\mathbf{z}+\\ \int_{0}^{\infty}\frac{e^{-(\mathbf{z}-\mathbf{\hat{p}})^2/2\boldsymbol{\tau}^p}}{\sqrt{2\pi \boldsymbol{\tau}^p}}((1-\epsilon)\delta(\mathbf{y}-1)+\epsilon\delta(\mathbf{y}+1))d\mathbf{z}
\end{multline}
If we calculate the integrals we obtain \(\mathbf{\hat{z}}^0\) as:
\begin{multline}
\mathbf{\hat{z}}^0=\frac{1}{\mathcal{Z}}(\frac{-e^{-\mathbf{\hat{p}}^2/2\boldsymbol{\tau}^p}\sqrt{\boldsymbol{\tau}^p}}{\sqrt{2\pi}}+\mathbf{\hat{p}}\frac{erfc(\mathbf{\hat{p}}/\sqrt{2\boldsymbol{\tau}^p})}{2})((1-\epsilon)\delta(\mathbf{y}+1)+\epsilon\delta(\mathbf{y}-1))+\\ \frac{1}{\mathcal{Z}}(\frac{e^{-\mathbf{\hat{p}}^2/2\boldsymbol{\tau}^p}\sqrt{\boldsymbol{\tau}^p}}{\sqrt{2\pi}}+\mathbf{\hat{p}}\frac{1+erf(\mathbf{\hat{p}}/\sqrt{2\boldsymbol{\tau}^p})}{2})((1-\epsilon)\delta(\mathbf{y}-1)+\epsilon\delta(\mathbf{y}+1))
\end{multline}
where \(\mathcal{Z}\) is:
\begin{multline}
\mathcal{Z}=\frac{1}{2}erfc(\mathbf{\hat{p}}/\sqrt{2\boldsymbol{\tau}^p})((1-\epsilon)\delta(\mathbf{y}+1)+\epsilon\delta(\mathbf{y}-1))+\\ \frac{1}{2}(1+erf(\mathbf{\hat{p}}/\sqrt{2\boldsymbol{\tau}^p}))((1-\epsilon)\delta(\mathbf{y}-1)+\epsilon\delta(\mathbf{y}+1))
\end{multline}
We use \(\mathbf{\hat{z}}^0\) to compute \(\mathbf{\hat{s}}\) which is 
\begin{equation}
\mathbf{\hat{s}}=\frac{\mathbf{\hat{z}}^0-\mathbf{\hat{p}}}{\boldsymbol{\tau}^p}
\end{equation}

After finding an expression for \(\mathbf{\hat{s}}\), we want to compute \(\boldsymbol{\tau}^s\) for BSC. To compute \(\boldsymbol{\tau}^{s}\), we need to find \(var(\mathbf{z}|\mathbf{\hat{p}},\mathbf{y},\boldsymbol{\tau}^p)\) which we can compute by:
\begin{equation}
var(\mathbf{z}|\mathbf{\hat{p}},\mathbf{y},\boldsymbol{\tau}^p)=\mathbb{E}(\mathbf{z}^2|\mathbf{\hat{p}},\mathbf{y},\boldsymbol{\tau}^p)-\mathbb{E}(\mathbf{z}|\mathbf{\hat{p}},\mathbf{y},\boldsymbol{\tau}^p)^2
\end{equation}
We know \(\mathbb{E}(\mathbf{z}|\mathbf{\hat{p}},\mathbf{y},\boldsymbol{\tau}^p)=\mathbf{\hat{z}}^0\) from the previous derivations, therefore we need to compute \(\mathbb{E}(\mathbf{z}^2)\) to compute \(var(\mathbf{z}|\mathbf{\hat{p}},\mathbf{y},\boldsymbol{\tau}^p)\):
\begin{multline}
\mathbb{E}(\mathbf{z}^2)=\frac{1}{\mathcal{Z}}\int_{-\infty}^{0}\frac{\mathbf{z}^2e^{-(\mathbf{z}-\mathbf{\hat{p}})^2/2\boldsymbol{\tau}^p}}{\sqrt{2\pi \boldsymbol{\tau}^p}}((1-\epsilon)\delta(\mathbf{y}+1)+\epsilon\delta(\mathbf{y}-1))d\mathbf{z}+\\ \frac{1}{\mathcal{Z}}\int_{0}^{\infty}\frac{\mathbf{z}^2e^{-(\mathbf{z}-\mathbf{\hat{p}})^2/2\boldsymbol{\tau}^p}}{\sqrt{2\pi \boldsymbol{\tau}^p}}((1-\epsilon)\delta(\mathbf{y}-1)+\epsilon\delta(\mathbf{y}+1))d\mathbf{z}
\end{multline}
After calculating the integrals, we obtain \(\mathbb{E}(\mathbf{z}^2)\) as 
\begin{multline}
 \mathbb{E}(\mathbf{z}^2)=\frac{1}{\mathcal{Z}}(\frac{-e^{-\mathbf{\hat{p}}^2/2\boldsymbol{\tau}^p}\mathbf{\hat{p}}\sqrt{\boldsymbol{\tau}^p}}{\sqrt{2\pi}}+(\mathbf{\hat{p}}^2+\boldsymbol{\tau}^p)\frac{erfc(\mathbf{\hat{p}}/\sqrt{2\boldsymbol{\tau}^p})}{2})((1-\epsilon)\delta(\mathbf{y}+1)+\epsilon\delta(\mathbf{y}-1))+\\ \frac{1}{\mathcal{Z}}(\frac{e^{-\mathbf{\hat{p}}^2/2\boldsymbol{\tau}^p}\mathbf{\hat{p}}\sqrt{\boldsymbol{\tau}^p}}{\sqrt{2\pi}}+(\mathbf{\hat{p}}^2+\boldsymbol{\tau}^p)\frac{1+erf(\mathbf{\hat{p}}/\sqrt{2\boldsymbol{\tau}^p})}{2})((1-\epsilon)\delta(\mathbf{y}-1)+\epsilon\delta(\mathbf{y}+1))
\end{multline}
Now we can compute \(var(\mathbf{z}|\mathbf{\hat{p}},\mathbf{y},\boldsymbol{\tau}^p)\) using the equation \(var(\mathbf{z}|\mathbf{\hat{p}},\mathbf{y},\boldsymbol{\tau}^p)=\mathbb{E}(\mathbf{z}^2)-(\mathbf{\hat{z}}^0)^2\). With the knowledge of \(var(\mathbf{z}|\mathbf{\hat{p}},\mathbf{y},\boldsymbol{\tau}^p)\), we can find an expression for \(\boldsymbol{\tau}^s\) which is:
\begin{equation}
\boldsymbol{\tau}^s=(\boldsymbol{\tau}^p-var(\mathbf{z}|\mathbf{\hat{p}},\mathbf{y},\boldsymbol{\tau}^p))/(\boldsymbol{\tau}^p)^2
\end{equation}

\subsubsection{Computing \(\mathbf{\hat{s}}\) and \(\boldsymbol{\tau}^s\) for Z Channel}
After finding \(\mathbf{\hat{s}}\) and \(\boldsymbol{\tau}^s\) for BSC, we can compute them for Z channel by replacing \(P_{Y|Z}\) by
\begin{equation}
P_{Y|Z}=\delta(\pi(\mathbf{z})-1)\delta(\mathbf{y}-\pi(\mathbf{z}))+\delta(\pi(\mathbf{z})+1)((1-\epsilon)\delta(\mathbf{y}-\pi(\mathbf{z}))+\epsilon\delta(\mathbf{y}+\pi(\mathbf{z})))
\end{equation}
where \(\epsilon\) is the flip probability of input -1. Now we replace the \(P_{Y|Z}\) term inside \(\mathbf{\hat{z}}^0\), \(\mathcal{Z}\) and \(\mathbb{E}(\mathbf{z})^2\) to find them for Z channel.
\begin{multline}
\mathbf{\hat{z}}^0=\frac{1}{\mathcal{Z}}(\frac{-e^{-\mathbf{\hat{p}}^2/2\boldsymbol{\tau}^p}\sqrt{\boldsymbol{\tau}^p}}{\sqrt{2\pi}}+\mathbf{\hat{p}}\frac{erfc(\mathbf{\hat{p}}/\sqrt{2\boldsymbol{\tau}^p})}{2})((1-\epsilon)\delta(\mathbf{y}+1))+\\ \frac{1}{\mathcal{Z}}(\frac{e^{-\mathbf{\hat{p}}^2/2\boldsymbol{\tau}^p}\sqrt{\boldsymbol{\tau}^p}}{\sqrt{2\pi}}+\mathbf{\hat{p}}\frac{1+erf(\mathbf{\hat{p}}/\sqrt{2\boldsymbol{\tau}^p})}{2})(\delta(\mathbf{y}-1)+\epsilon\delta(\mathbf{y}+1))
\end{multline}
\begin{multline}
\mathcal{Z}=\frac{1}{2}erfc(\mathbf{\hat{p}}/\sqrt{2\boldsymbol{\tau}^p})((1-\epsilon)\delta(\mathbf{y}+1))+\\ \frac{1}{2}(1+erf(\mathbf{\hat{p}}/\sqrt{2\boldsymbol{\tau}^p}))(\delta(\mathbf{y}-1)+\epsilon\delta(\mathbf{y}+1))
\end{multline}
\begin{multline}
 \mathbb{E}(\mathbf{z}^2)=\frac{1}{\mathcal{Z}}(\frac{-e^{-\mathbf{\hat{p}}^2/2\boldsymbol{\tau}^p}\mathbf{\hat{p}}\sqrt{\boldsymbol{\tau}^p}}{\sqrt{2\pi}}+(\mathbf{\hat{p}}^2+\boldsymbol{\tau}^p)\frac{erfc(\mathbf{\hat{p}}/\sqrt{2\boldsymbol{\tau}^p})}{2})((1-\epsilon)\delta(\mathbf{y}+1))+\\ \frac{1}{\mathcal{Z}}(\frac{e^{-\mathbf{\hat{p}}^2/2\boldsymbol{\tau}^p}\mathbf{\hat{p}}\sqrt{\boldsymbol{\tau}^p}}{\sqrt{2\pi}}+(\mathbf{\hat{p}}^2+\boldsymbol{\tau}^p)\frac{1+erf(\mathbf{\hat{p}}/\sqrt{2\boldsymbol{\tau}^p})}{2})(\delta(\mathbf{y}-1)+\epsilon\delta(\mathbf{y}+1))
\end{multline}
Now we can find \(\mathbf{\hat{s}}\) and \(\boldsymbol{\tau}^s\) using the \(\mathbf{\hat{z}^0}\), \(\mathcal{Z}\) and \(\mathbb{E}(\mathbf{z})^2\) expressions for Z channel by using the formulas \(\mathbf{\hat{s}}=\frac{\mathbf{\hat{z}}^0-\mathbf{\hat{p}}}{\boldsymbol{\tau}^p}\) and \(\boldsymbol{\tau}^s=(\boldsymbol{\tau}^p-var(\mathbf{z}|\mathbf{\hat{p}},\mathbf{y},\boldsymbol{\tau}^p))/(\boldsymbol{\tau}^p)^2\) where \(var(\mathbf{z}|\mathbf{\hat{p}},\mathbf{y},\boldsymbol{\tau}^p)=\mathbb{E}(\mathbf{z}^2)-(\mathbf{\hat{z}}^0)^2\).

\section{State Evolution Function}

In order to evaluate the performance of our decoder, we need an objective measure to compare it. Thus, BSC and Z Channel are modeled as AWGN Channels with an effective noise variance \(\Sigma(E)^{2}\). Then, we iteratively compute the mean square error (MSE) \(\widetilde{E}^{(t)}=\mathbb{E}_{\mathbf{x,y}}[\frac{1}{L}\sum_{n=1}^{N}(\hat{x}^{(t)}_{n}-x_{n})^2]\) of the GAMP estimate (\(\mathbf{\hat{x}}^{(t)}\) in our case) for every iteration. 

To do it, firstly, we calculated the effective noise variance \(\Sigma(E)^{2}\) by the relation:
\begin{equation}
\Sigma(E)^{2}=\frac{R}{\mathbb{E}_{p|E}[\mathcal{F}(p|E)] }
\end{equation}

where the expectation \(\mathbb{E}_{p|E}\) is with respect to \(\mathcal{N}(p|0,1-E)\) and
\begin{equation}
\mathcal{F}(p|E)=\int  \mathrm{d}y  f(y|p,E) (\partial_{m} \ln f(y|m,E))^{2}_{m=p}
\end{equation}

is the Fisher information of \(p\) associated with the distribution below:
\begin{equation}
f(y|p,E)=\int\mathrm{d}u P_{out}(y|u)\mathcal{N}(u|p,E)
\end{equation}

The calculation of the Fisher information is given for BSC below:
\begin{equation}
\begin{multlined}
f(y|p,E)=\int_{-\infty}^{0} \mathrm{d}u [(1-\epsilon)\delta(y+1)+\epsilon\delta(y-1)]\frac{e^{\frac{-(u-p)^{2}}{2E}}}{\sqrt {2\pi E}}\\ \\
+\int_{0}^{\infty} \mathrm{d}u [(1-\epsilon)\delta(y-1)+\epsilon\delta(y+1)]\frac{e^{\frac{-(u-p)^{2}}{2E}}}{\sqrt {2\pi E}}
\end{multlined}
\end{equation}

By change of variables, we write \(t=\frac{u-p}{\sqrt{E}}\), which then gives us:
\begin{equation}
\begin{multlined}
f(y|p,E)=\int_{-\infty}^{\frac{-p}{\sqrt{E}}} \mathrm{d}t [(1-\epsilon)\delta(y+1)+\epsilon\delta(y-1)]\frac{e^{\frac{-t^{2}}{2}}}{\sqrt {2\pi}}\\ \\
+\int_{\frac{-p}{\sqrt{E}}}^{\infty} \mathrm{d}t [(1-\epsilon)\delta(y-1)+\epsilon\delta(y+1)]\frac{e^{\frac{-t^{2}}{2}}}{\sqrt {2\pi}}
\end{multlined}
\end{equation}

As one can see, these integrals are in the form of \(Q\) functions, therefore we finally declare \(f(y|p,E)\) as:
\begin{equation}
\begin{multlined}
f(y|p,E)= (1-\epsilon)[\delta(y-1)\mathbb{Q}(\textstyle \frac{-p}{\sqrt{E}})+\delta(y+1)(1-\mathbb{Q}(\frac{-p}{\sqrt{E}}))]\\
+ \epsilon[\delta(y+1)\mathbb{Q}(\textstyle\frac{-p}{\sqrt{E}})+\delta(y-1)(1-\mathbb{Q}(\frac{-p}{\sqrt{E}}))]
\end{multlined}
\end{equation}

Now, we can calculate \(\mathcal{F}(p|E)\), since we have 
\(f(y|p,E)\). From the definition of \(\mathcal{F}(p|E)\), it is equivalent to:
\begin{equation}
\mathcal{F}(p|E)=\int  \mathrm{d}y  \frac {(\partial_{p} f(y|p,E))^{2}} {f(y|p,E)}
\end{equation}

Let us say \(Q=\mathbb{Q}(\textstyle\frac{-p}{\sqrt{E}}) \) 
for the \(Q\) functions (\(Q'\) to their derivatives with respect to \(p\)) which take place in \(f(y|p,E)\). We do this change for the ease of understanding of the calculation for the following steps:
\begin{equation}
\mathcal{F}(p|E)=\int  \mathrm{d}y  \frac  {[(1-\epsilon)(\delta(y-1)Q'-\delta(y+1)Q')+\epsilon(\delta(y+1)Q'-\delta(y-1)Q')]^{2}} {(1-\epsilon)[\delta(y-1)Q+\delta(y+1)(1-Q)]+\epsilon[\delta(y+1)Q+\delta(y-1)(1-Q)]}
\end{equation}

Since \(\mathbf{y}\) is the output vector from the binary symmetric channel, it only consists of  \(-1,1\)'s. Due to this fact, integration can be done by first making \(y=-1\), then \(y=1\) and finally summing these two results. Combining these facts, we finally get:
\begin{equation}
\mathcal{F}(p|E)=\frac{(Q'-2\epsilon Q')^{2}}{(Q+\epsilon-2\epsilon Q)(1-Q-\epsilon+2\epsilon Q)}
\end{equation}

where 
\begin{equation}
Q=\int_{\frac{-p}{\sqrt{E}}}^{\infty} \mathrm{d}t \frac
{e^{\frac{-t^{2}}{2}}}{\sqrt{2\pi}}   \qquad
and \qquad
Q'=\frac{e^{\frac{-p^{2}}{2E}}}{\sqrt{2\pi E}}
\end{equation}

As a result, Fisher information \(\mathcal{F}(p|E)\) calculation is completed. It can also be computed for the Z Channel in the same way just by changing \(P_{out}(y|u)\) from \( (1-\epsilon)(\delta(y-\pi(u))+\epsilon\delta(y+\pi(u))\) to \( \delta(\pi(u)-1)\delta(y-\pi(u))+\delta(\pi(u)+1)[(1-\epsilon)(\delta(y-\pi(u))+\epsilon\delta(y+\pi(u))\) where \(\pi(u)=sign(u)\). 

After Fisher information calculation step, \(\mathbb{E}_{p|E}[\mathcal{F}(p|E)]\) can be found by taking the expectation of \(\mathcal{F}(p|E)\) with respect to \(\mathcal{N}(p|0,1-E)\). This is done by Monte Carlo integration using Matlab. Using this result and the formula (12), effective noise variance is calculated which makes us ready to continue for the second part to complete the state evolution function.


At the previous part, we computed \(\Sigma(E)^{2}\) for BSC and Z channel and we can model these channels as an effective AWGN channel with noise \(\mathcal{N}(0,\Sigma(E)^{2}/log_{2}(B))\). Then we create a denoiser \(f_{i}(\Sigma(E))\) which estimates the value of the \(i\)-th component of a section. We can define \(f_{i}(\Sigma(E))\) as:
\begin{equation}
    f_{i}(\Sigma(E))=\frac{\Sigma_{\mathbf{x}}e^{-\frac{\|x-(s+z\Sigma(E)/\sqrt{log_{2}(B)}\|_{2}^{2}}{2\Sigma(E)^{2}/log_{2}(B)}}p_{0}(\mathbf{x})x_{i}}{\Sigma_{\mathbf{x}}e^{-\frac{\|x-(s+z\Sigma(E)/\sqrt{log_{2}(B)}\|_{2}^{2}}{2\Sigma(E)^{2}/log_{2}(B)}}p_{0}(\mathbf{x})}
\end{equation}
where \(s_{i}\) is the correct value of the \(i\)-th component of a section and \(z\) is a random variable with distribution \(\mathcal{N}(0,1)\). We can simplify that equation such that:
\begin{equation}
 f_{i}(\Sigma(E))=[1+\Sigma_{k\ne i}^{B}e^{(s_{k}-s_{i})log_{2}(B)/\Sigma(E)^2+(z_{k}-z_{i})\sqrt{log_{2}(B)}/\Sigma(E)}]^{-1}
\end{equation}
Therefore MSE estimated from state evolution becomes:
\begin{equation}
    T_{u}(E)=\mathbb{E}_{s,z}[\Sigma_{i=1}^B(f_{i}(\Sigma(E))-s_{i})^2]
\end{equation}
We can simplify that equation further, because we know that only one of the \(s_{i}\) values in a section is 1 while the other \((B-1)\) of them are 0s. Lets assume that \(s_{1}=1\) while other \(s_{i}=0\) \(\forall i\in\{2,...,B\}\). We know that \(f_{i}\) for \(\forall i\in\{2,...,B\}\) has an identical distribution because \(s_{i}=0\) and \(z_{i}\) is a random variable with distribution \(\mathcal{N}(0,1)\). Therefore we can compute the distribution of \(s_{2}\) and assign that value for \(\forall i\in\{3,...,B\}\).
\begin{equation}
    f_{1}(\Sigma(E))=[1+e^{-log_{2}(B)/\Sigma(E)^2}\Sigma_{k\ge 2}^{B}e^{(z_{k}-z_{1})\sqrt{log_{2}(B)}/\Sigma(E)}]^ {-1}
\end{equation}
\begin{equation}
    f_{2}(\Sigma(E))=[1+e^{log_{2}(B)/\Sigma(E)^2+(z_{1}-z_{2})\sqrt{log_{2}(B)}/\Sigma(E)}+\Sigma_{k\ge 3}^{B}e^{(z_{k}-z_{2})\sqrt{log_{2}(B)}/\Sigma(E)}]^ {-1}
\end{equation}
Therefore \(T_{u}(E)\) becomes,
\begin{equation}
    T_{u}(E)=\int \mathcal{D} z (f_{1}(\Sigma(E))-1)^2+(B-1)f_{2}(\Sigma(E))^2
\end{equation}
To compute this integral, we can use Monte Carlo integration because \(z\) is a random variable with a distribution \(\mathcal{N}(0,1)\). Therefore, we can give z multiple random values with distribution \(\mathcal{N}(0,1)\) and take the mean of the results to obtain \(T_{u}(E)\).

Section error rate is another way to evaluate the performance of the GAMP decoder and we can also estimate the value of SER using state evolution. We obtained the equations for \(f_{1}(\Sigma(E))\) for \(s_{1}=1\) and \(f_{i}(\Sigma(E))\) for \(s_{i}=0\), \(\forall i\in\{2,...,B\}\). We know that \(f_{i}(\Sigma(E))\) estimates \(\hat{s_{i}}\) values, therefore we can estimate SER by using:
\begin{equation}
    SER=\int \mathcal{D} z  \mathbb{1}(\exists i\in\{2,...,B\} : f_{i}(\Sigma(E))>f_{1}(\Sigma(E)))
\end{equation}
We can compute this integral using Monte Carlo integration the way we computed  \(T_{u}(E)\).

\section{Potential Function for AWGN Channel}
    Another way to determine the performance of our decoder is to evaluate its potential function. For AWGN channel we can define the potential function as:
    \begin{multline}
        \Phi_{B}(E)=-\frac{\log_2(B)}{2R}\left(log(1/snr+E)+\frac{1-E}{1/snr+E}\right)+\\ \int \mathcal{D}\bar{z}log\left(e^{\frac{1}{2\Sigma(E)^2}+\frac{z_1}{\Sigma(E)}}+\Sigma_{i=2}^{B}e^{-\frac{1}{2\Sigma(E)^2}+\frac{z_1}{\Sigma(E)}}\right)
    \end{multline}
where \(E=\frac{1}{L}\sum_{i}^{N}(x_{i}-\hat{x_{i}})^2\), \(\Sigma(E):=\sqrt{(\frac{1}{snr}+E)R/log_{2}B}\) and \(\mathcal{D}\bar{z}=\Pi_{i=1}^B\mathcal{D}z_{i}\) where \(\mathcal{D}z_{i}\) has the distribution of \(\mathcal{N}(0,1)\). To take the integral we can use Monte Carlo integration.

The information we gain from the potential function is the behavior of \(E\) for different \(R\),\(B\) and \(snr\). To observe the behaviour of \(E\), we need to search for the local maxima of the potential function by decreasing \(E\). By looking at the \(E\) value at the local maxima, we can observe the converging \(E\) value of the decoder.

What we can say about the shape of the potential function graph is that, if the potential function only has one local maxima, then the corresponding \(E\) is very small. However, if the potential function stays constant if we decrease \(E\) and starts to increase for smaller \(E\), we are at the belief propagation threshold and a small decrease in \(R\) may create a local maxima at that point, which may increase \(E\) by a big amount. Another scenario is to have two global maximas for different \(E\) values and the one with the higher error dominates the other one. We call that rate as the optimal threshold, which is the highest rate with a bad performance under optimal decoding. 

\begin{figure}[t!]
\centering
\begin{minipage}{.5\textwidth}
  \centering
  \includegraphics[width=1\linewidth]{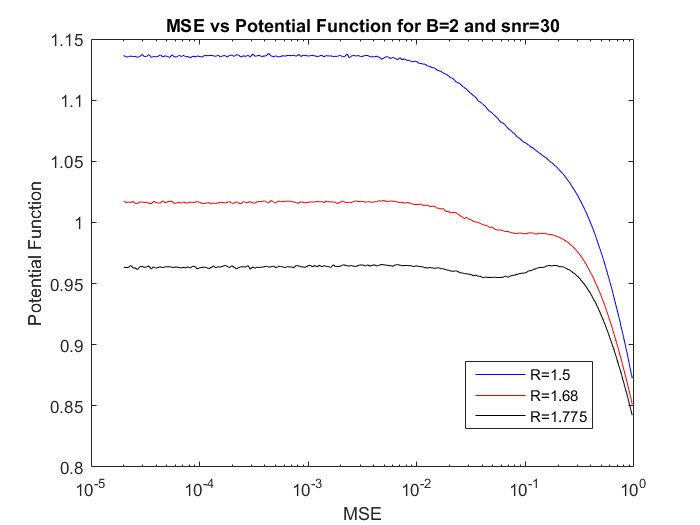}
  \captionof{figure}{MSE vs Potential Function for B=2 and snr=30}
  \label{fig:pot1}
\end{minipage}%
\begin{minipage}{.5\textwidth}
  \centering
  \includegraphics[width=1\linewidth]{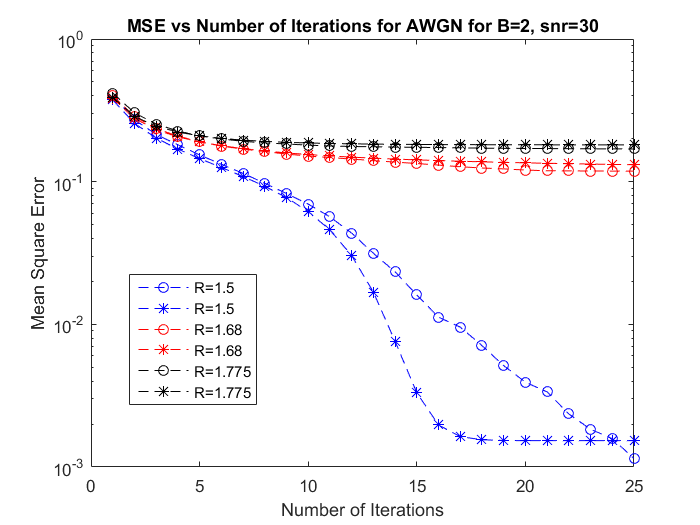}
  \captionof{figure}{Iteartion vs MSE for B=2 and snr=30}
  \label{fig:pot2}
\end{minipage}
\end{figure}

To observe the behaviour of the potential function, we selected our \(B=2\), \(snr=30\) and \(R=1.5, 1.68, 1.775\). From Figure 1, we can see that \(R=1.775\) is the optimal threshold because there are 2 global maximas at the potential function. Also, we can say that \(R=1.68\) is at the belief propagation threshold, because the potential function stays constant for a while then increases when \(E\) decreases. \(R=1.5\) is below the belief propagation threshold, because it only has one global maxima.

In Figure 1, when \(R=1.775\), the local maxima appears when \(E=0.1834\) while in Figure 2 \(E\) converges to \(0.1811\). So, we can estimate the converging \(E\) of the decoder by looking at the \(E\) value of the local maxima in the potential function when \(R\) is greater or equal to the optimal threshold. In Figure 1, when \(R=1.68\), the local maxima appears when \(E=0.1509\) while in Figure 2 \(E\) converges to \(0.1314\). Therefore, we can use \(E\) of the local maxima at the potential function to determine the \(E\) of the decoder. However, it is not as precise as when \(R\) is greater or equal to the optimal threshold, because at BP threshold region, small changes on \(R\) changes the \(E\) of the local maxima of the potential function and \(E\) obtained at the decoder by a huge amount. In Figure 1, when \(R=1.5\), the local maxima appears when \(E=0.0098\) while in Figure 2 \(E\) converges to \(0.0015\). We can still use the information about \(E\) of the local maxima of the potential function, but as \(E\) decreases below \(10^{-2}\), potential function graph stays constant, therefore we need to have extremely high number of z values in our Monte Carlo integration to determine the exact location of the local maxima.

\section{Experimental Results and Discussion}

We now present a number of numerical experiments testing the performance and behavior of the GAMP decoder in different practical scenarios with finite size signals. Communication rate \(R\), section length \(B\), flip probability \(\epsilon\), signal-to-noise ratio \(snr\) and the channel used in communication are varied in these experiments. To evaluate the results, we plotted many observations. 

\begin{figure}[h!]
\centering
\begin{minipage}{.5\textwidth}
  \centering
  \includegraphics[width=1\linewidth]{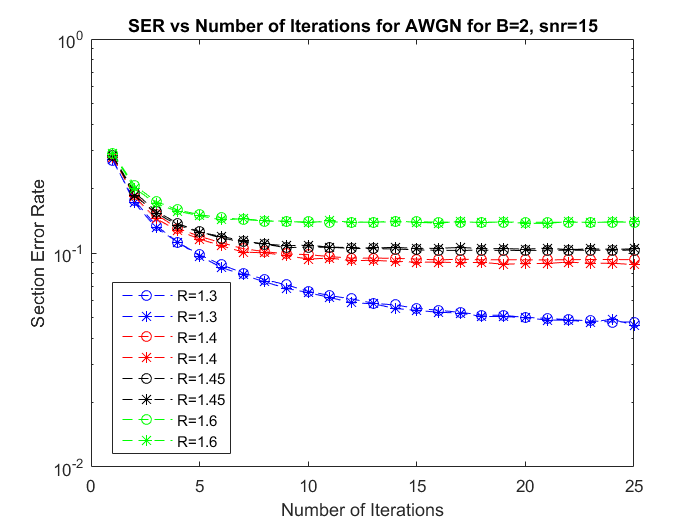}
  \captionof{figure}{Iteration vs SER in AWGN \newline(\(snr=15\) , \(B=2\))}
  \label{fig:pot1}
\end{minipage}%
\begin{minipage}{.5\textwidth}
  \centering
  \includegraphics[width=1\linewidth]{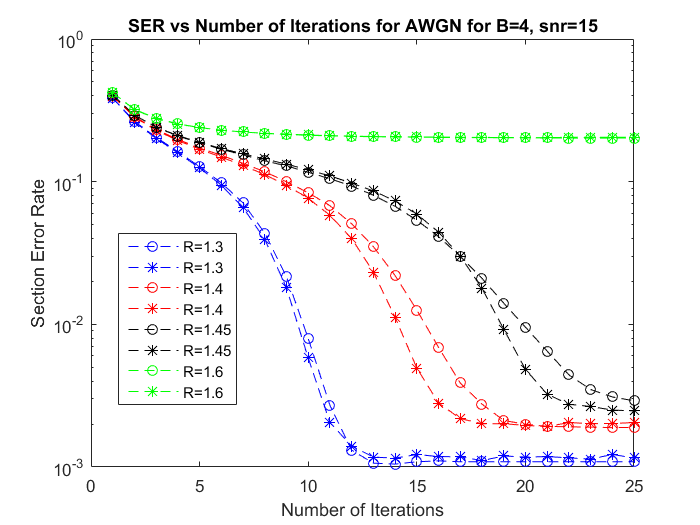}
  \captionof{figure}{Iteration vs SER in AWGN \newline( \(snr=15\) , \(B=4\))}
  \label{fig:pot2}
\end{minipage}
\end{figure}

First of all, iteration vs symbol error rate (SER) plots for AWGN Channel where \(B=2, 4\) and \(SNR=15\) can be seen in Figures 3, 4. In the figures, 'o' denotes the decoder's SER and '*' denotes the state evolution's SER. The results of the decoder is followed by the state evolution function as expected. Therefore, the decoder works in AWGN Channel. It is also possible to confirm from the figures that when the communication rate increases it becomes harder to decode the signal.
\begin{figure}[b!]
\centering
\begin{minipage}{.5\textwidth}
  \centering
  \includegraphics[width=1\linewidth]{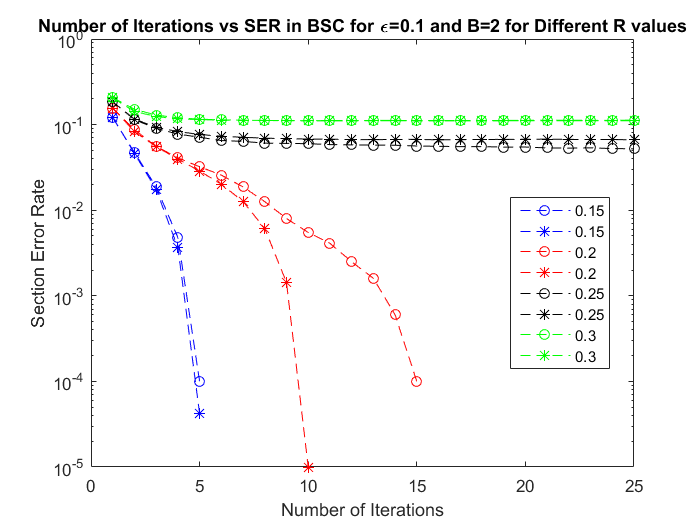}
  \captionof{figure}{Iteration vs SER in BSC (\(\epsilon=0.1\) , \(B=2\))}
  \label{fig:pot1}
\end{minipage}%
\begin{minipage}{.5\textwidth}
  \centering
  \includegraphics[width=1\linewidth]{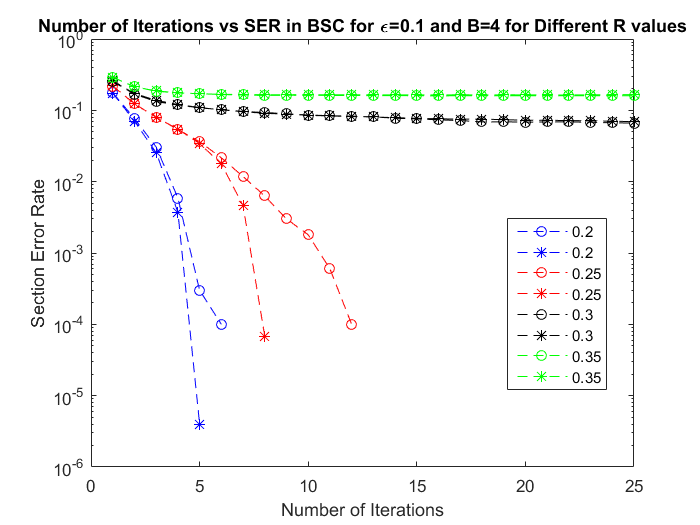}
  \captionof{figure}{Iteration vs SER in BSC ( \(\epsilon=0.1\) , \(B=4\))}
  \label{fig:pot2}
\end{minipage}
\end{figure}

\begin{figure}[t!]
\centering
\begin{minipage}{.5\textwidth}
  \centering
  \includegraphics[width=1\linewidth]{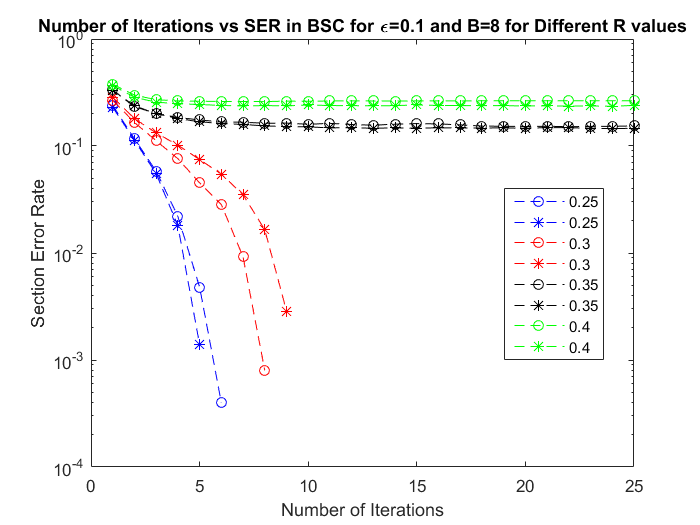}
  \captionof{figure}{Iteration vs SER in BSC (\(\epsilon=0.1\) , \(B=8\))}
  \label{fig:pot1}
\end{minipage}%
\begin{minipage}{.5\textwidth}
  \centering
  \includegraphics[width=1\linewidth]{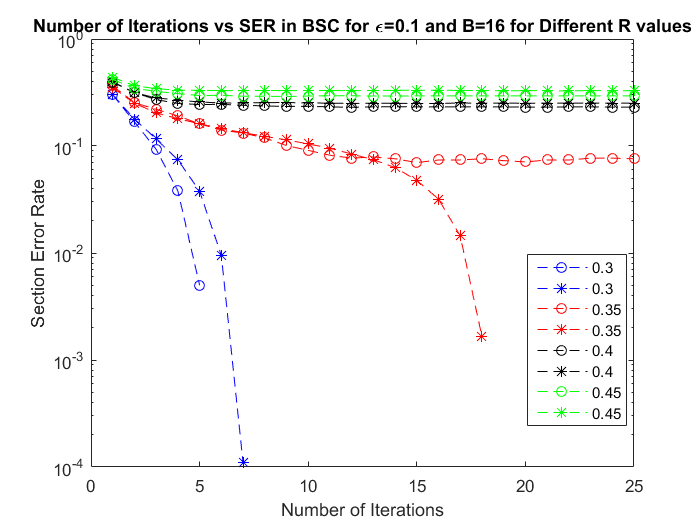}
  \captionof{figure}{Iteration vs SER in BSC (\(\epsilon=0.1\) , \(B=16\))}
  \label{fig:pot2}
\end{minipage}
\end{figure}

Secondly, iteration vs symbol error rate (SER) plots for BSC where \(B=2, 4, 8, 16\) and \(\epsilon=0.1\) can be seen in Figures 5, 6, 7 and 8. As seen, the results of the decoder and the state evolution function are coherent which proves that the decoder works well. 

Also, we should say that there are some differences like the one in Figure 8 where the communication rate equals \(R=0.35\), and this is stemmed from the fact that decoder is working near the transition region, and because we are working with finite size signals, those kind of variations can be observed from simulation to simulation.

\begin{figure}[h!]
\centering
\begin{minipage}{.5\textwidth}
  \centering
  \includegraphics[width=1\linewidth]{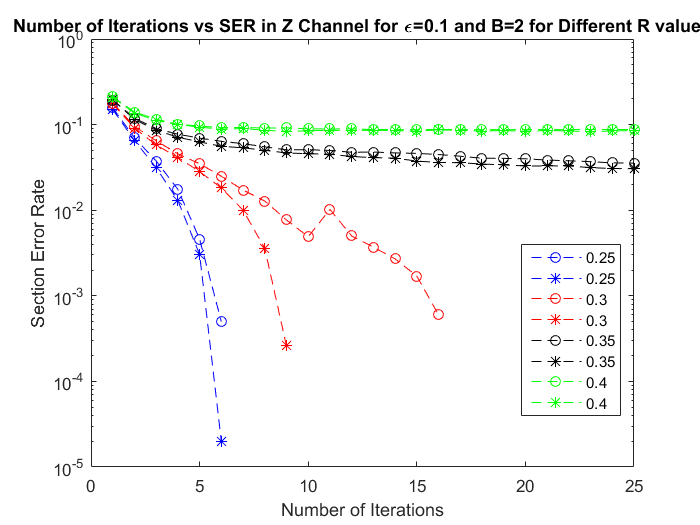}
  \captionof{figure}{Iteration vs SER in Z Channel \newline(\(\epsilon=0.1\) , \(B=2\))}
  \label{fig:pot1}
\end{minipage}%
\begin{minipage}{.5\textwidth}
  \centering
  \includegraphics[width=1\linewidth]{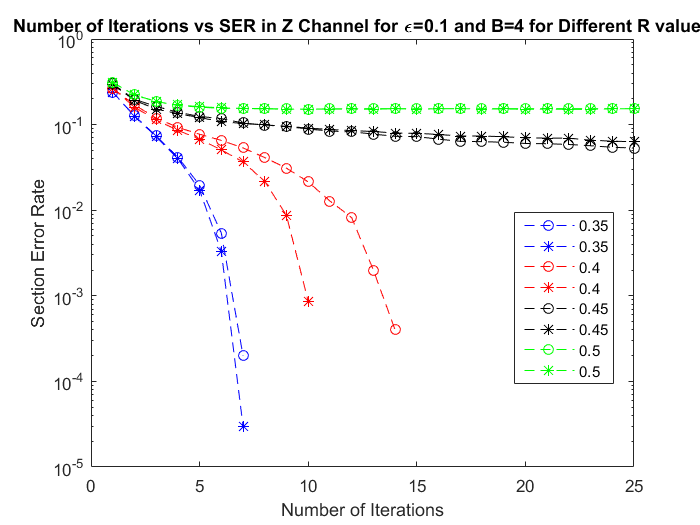}
  \captionof{figure}{Iteration vs SER in Z Channel \newline(\(\epsilon=0.1\) , \(B=4\))}
  \label{fig:pot2}
\end{minipage}
\end{figure}

\begin{figure}[h!]
\centering
\begin{minipage}{.5\textwidth}
  \centering
  \includegraphics[width=1\linewidth]{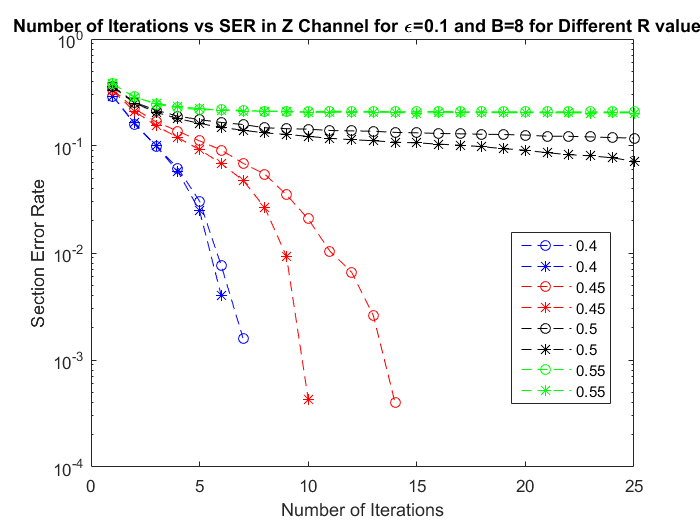}
  \captionof{figure}{Iteration vs SER in Z Channel \newline(\(\epsilon=0.1\) , \(B=8\))}
  \label{fig:pot1}
\end{minipage}%
\begin{minipage}{.5\textwidth}
  \centering
  \includegraphics[width=1\linewidth]{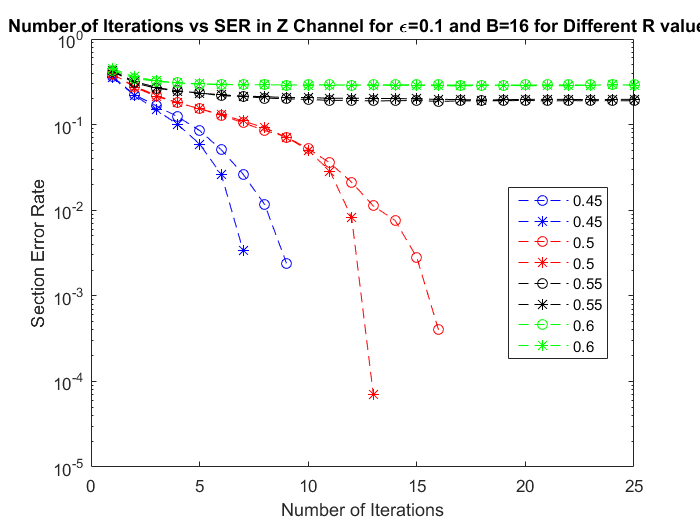}
  \captionof{figure}{Iteration vs SER in Z Channel \newline(\(\epsilon=0.1\) , \(B=16\))}
  \label{fig:pot2}
\end{minipage}
\end{figure}

In third scenario, iteration vs symbol error rate (SER) plots for Z Channel where \(B=2, 4, 8, 16\) and \(\epsilon=0.1\) can be seen in Figures 9, 10, 11 and 12. The correctness of the results is again confirmed by the state evolution function. From the figures, it is obvious that it becomes easier to decode the original signal if the communication occurs in lower rates.

In Figures 13, 15 and 17, we can see the convergence rates(\(R_{u}\)) of our decoder(the rate where SER becomes 0) for BSC for \(\epsilon=(0.01,0.05,0.1)\). In Figures 14, 16 and 18, we can see the convergence rates of our decoder for Z channel for \(\epsilon=(0.01,0.05,0.1)\). In each plot we show the convergence region for \(B=2, 4, 8, 16\). By comparing the figures, we can see that \(R_{u}\) increases when \(\epsilon\) decreases if we keep all other parameters constant. Also at the convergence region, we can see that the SER estimated from state evolution decreases sharply, while the SER of the decoder decreases more slowly. This effect happens because our signal is a finite size signal and if we increase L, the decrease of the SER around the convergence region becomes sharper.

\begin{figure}[h!]
\centering
\begin{minipage}{.5\textwidth}
  \centering
  \includegraphics[width=1\linewidth]{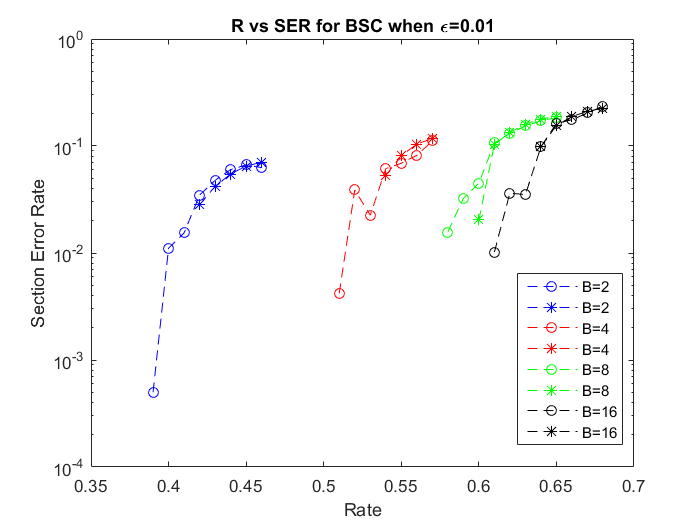}
  \captionof{figure}{Rate vs SER in BSC (\(\epsilon=0.01\))}
  \label{fig:pot1}
\end{minipage}%
\begin{minipage}{.5\textwidth}
  \centering
  \includegraphics[width=1\linewidth]{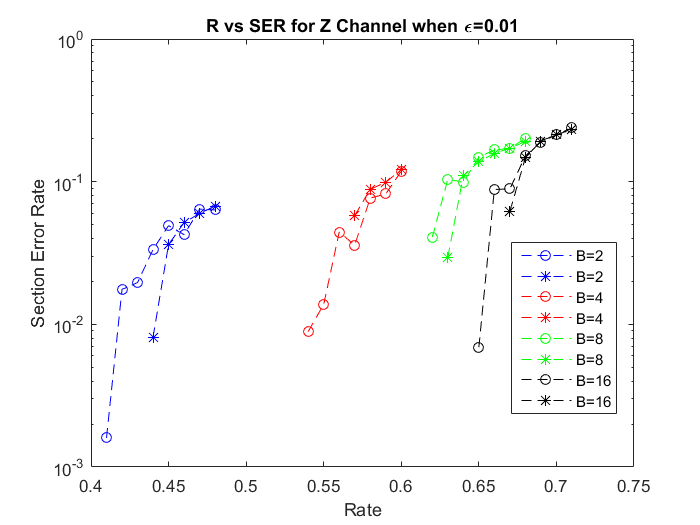}
  \captionof{figure}{Rate vs SER in Z Channel (\(\epsilon=0.01\))}
  
  \label{fig:pot2}
\end{minipage}
\end{figure}

\begin{figure}[h!]
\centering
\begin{minipage}{.5\textwidth}
  \centering
  \includegraphics[width=1\linewidth]{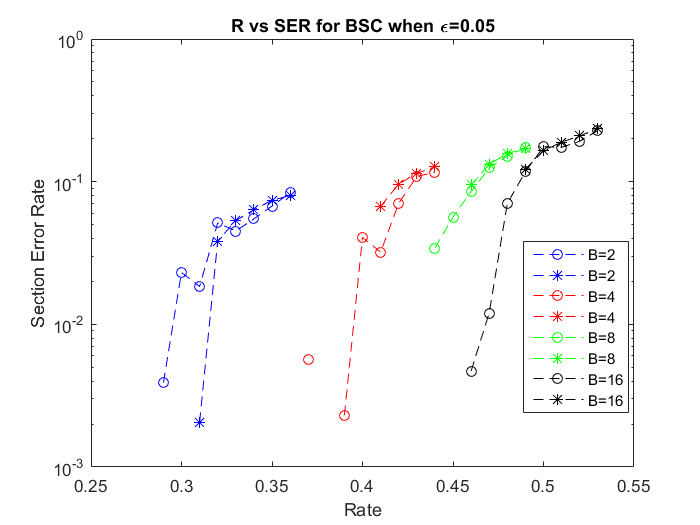}
  \captionof{figure}{Rate vs SER in BSC (\(\epsilon=0.05\))}
  \label{fig:pot1}
\end{minipage}%
\begin{minipage}{.5\textwidth}
  \centering
  \includegraphics[width=1\linewidth]{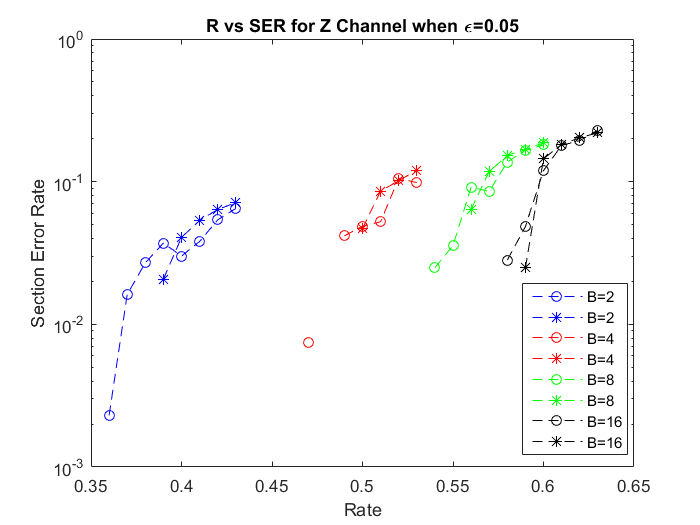}
  \captionof{figure}{Rate vs SER in Z Channel (\(\epsilon=0.05\))}
  \label{fig:pot2}
\end{minipage}
\end{figure}

\begin{figure}[h!]
\centering
\begin{minipage}{.5\textwidth}
  \centering
  \includegraphics[width=1\linewidth]{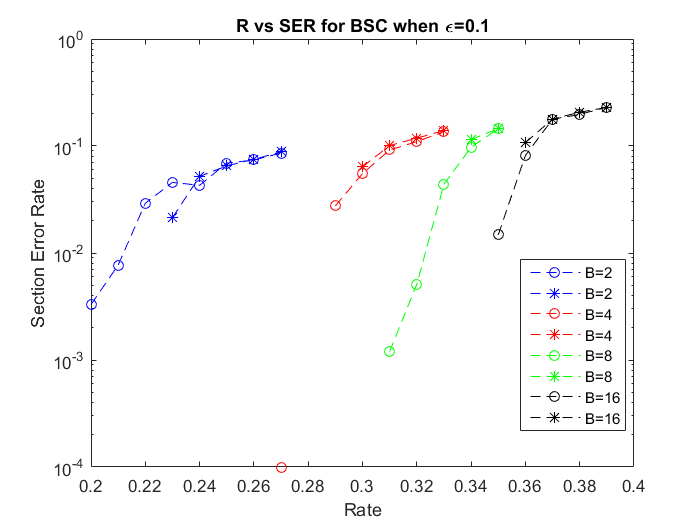}
  \captionof{figure}{Rate vs SER in BSC (\(\epsilon=0.1\))}
  \label{fig:pot1}
\end{minipage}%
\begin{minipage}{.5\textwidth}
  \centering
  \includegraphics[width=1\linewidth]{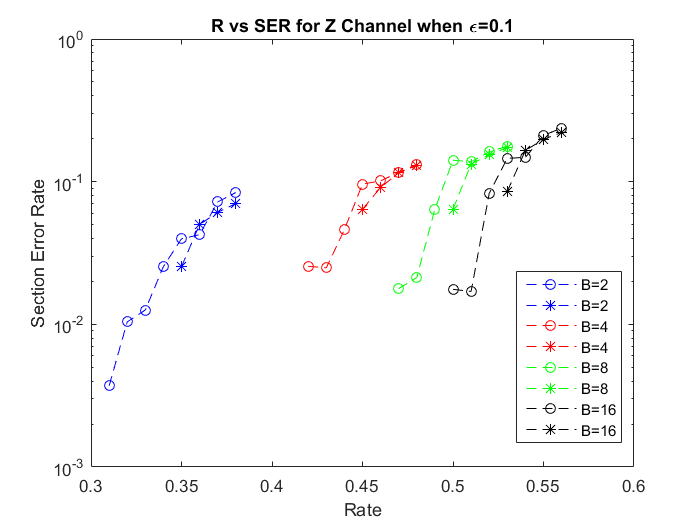}
  \captionof{figure}{Rate vs SER in Z Channel (\(\epsilon=0.1\))}
  \label{fig:pot2}
\end{minipage}
\end{figure}

For BSC and Z Channel, it is observed that when the same value of the communication rate \(R\) is used for different section length \(B's\), higher \(B\) gives lower section error rate. Besides, we can say that, when \(B\) increases \(R_{u}\) also increases if we keep all other parameters same.

We took the plots of the \(R_{u}^B(B)\) where \(B=2, 4, 8, 16, 32, 64\) and \(\epsilon=0.01, 0.05, 0.1\) for BSC and Z Channel. The plots can be seen in Figures 19, 20, 21, 22, 23, 24. To determine the correct \(R_{u}^B(B)\) values, we created an algorithm which has initial \(R_{min}\) and \(R_{max}\) values and we compute SER for \(R=(R_{min}+R_{max})/2\) and we repeat it for a huge number of times. If the decoder has \(SER=0\) for more than half of the runs, we set \(R_{min}=R\) and otherwise we set \(R_{max}=R\) and re-run the algorithm again. If \(R_{min}-R_{max}<T\), we stop the algorithm and select \(R_{u}^B(B)=(R_{min}+R_{max})/2\) for a given value of B and a threshold T.

\begin{figure}[h!]
\centering
\begin{minipage}{.5\textwidth}
  \centering
  \includegraphics[width=1\linewidth]{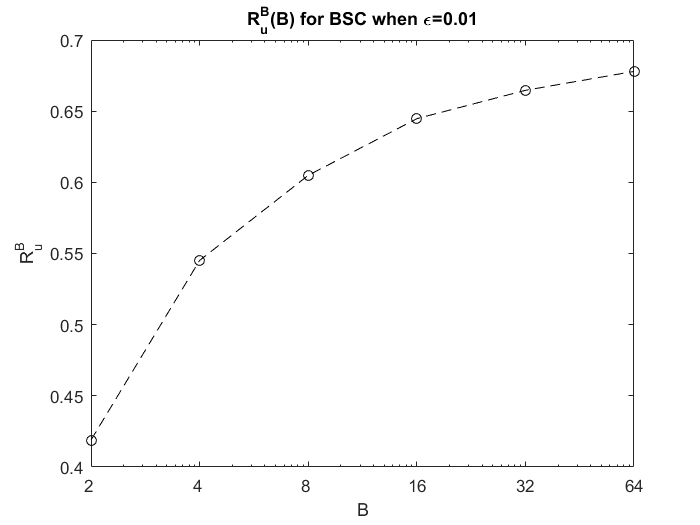}
  \captionof{figure}{\(B\) vs \(R_{u}\) in BSC (\(\epsilon=0.01\))}
  \label{fig:pot1}
\end{minipage}%
\begin{minipage}{.5\textwidth}
  \centering
  \includegraphics[width=1\linewidth]{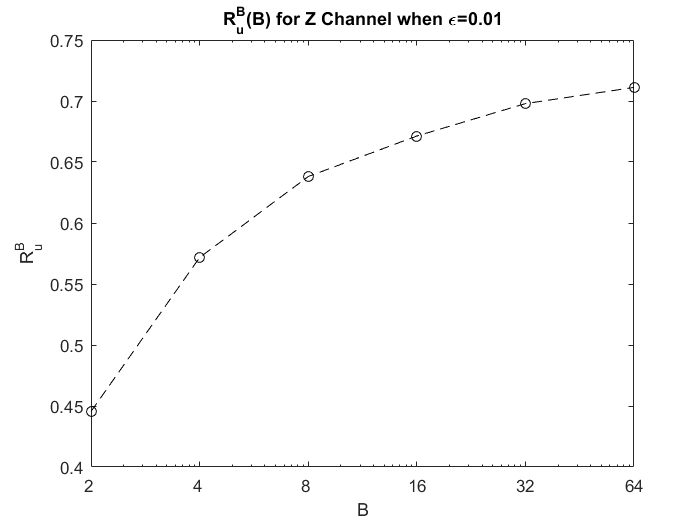}
  \captionof{figure}{\(B\) vs \(R_{u}\) in Z Channel (\(\epsilon=0.01\))}
  \label{fig:pot2}
\end{minipage}
\end{figure}

\begin{figure}[h!]
\centering
\begin{minipage}{.5\textwidth}
  \centering
  \includegraphics[width=1\linewidth]{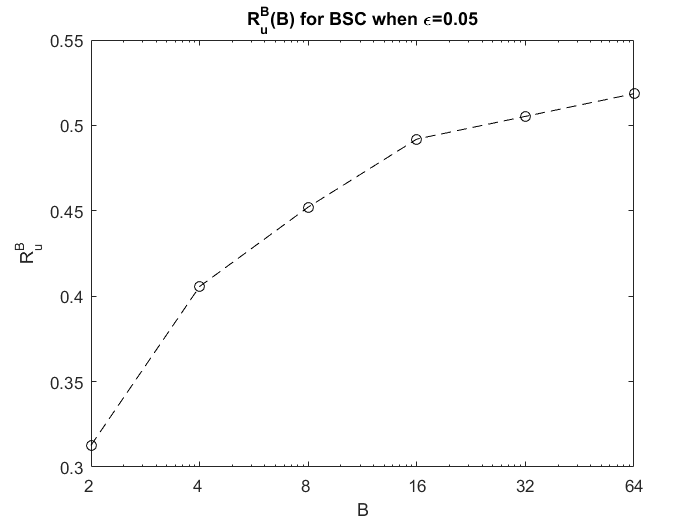}
  \captionof{figure}{\(B\) vs \(R_{u}\) in BSC (\(\epsilon=0.05\))}
  \label{fig:pot1}
\end{minipage}%
\begin{minipage}{.5\textwidth}
  \centering
  \includegraphics[width=1\linewidth]{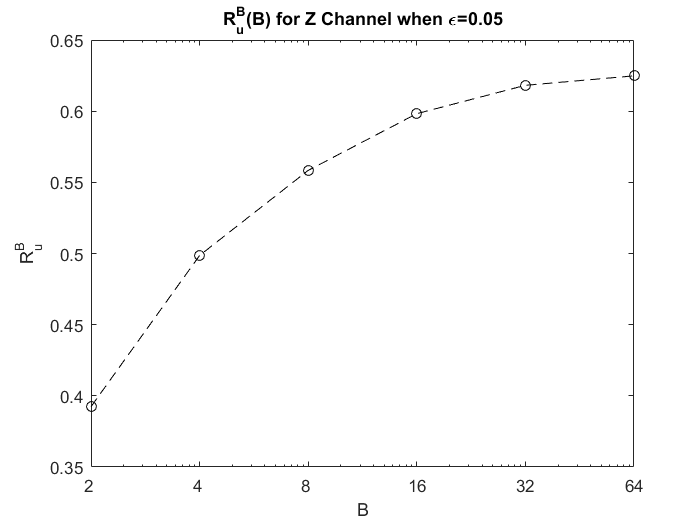}
  \captionof{figure}{\(B\) vs \(R_{u}\) in Z Channel (\(\epsilon=0.05\))}
  \label{fig:pot2}
\end{minipage}
\end{figure}

\begin{figure}[h!]
\centering
\begin{minipage}{.5\textwidth}
  \centering
  \includegraphics[width=1\linewidth]{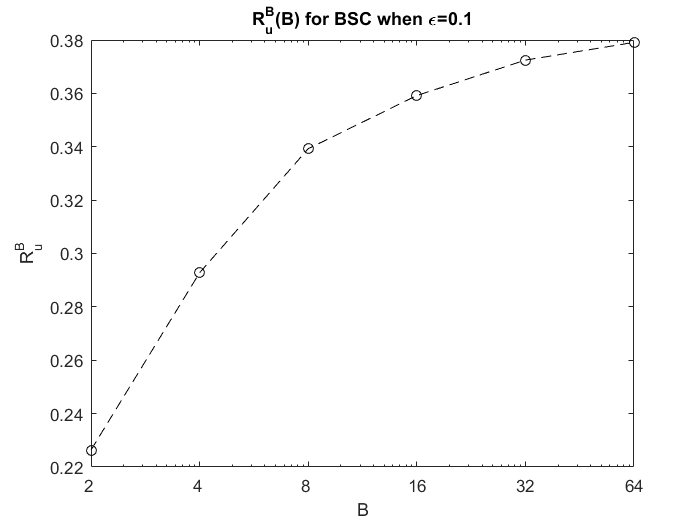}
  \captionof{figure}{\(B\) vs \(R_{u}\) in BSC (\(\epsilon=0.1\))}
  \label{fig:pot1}
\end{minipage}%
\begin{minipage}{.5\textwidth}
  \centering
  \includegraphics[width=1\linewidth]{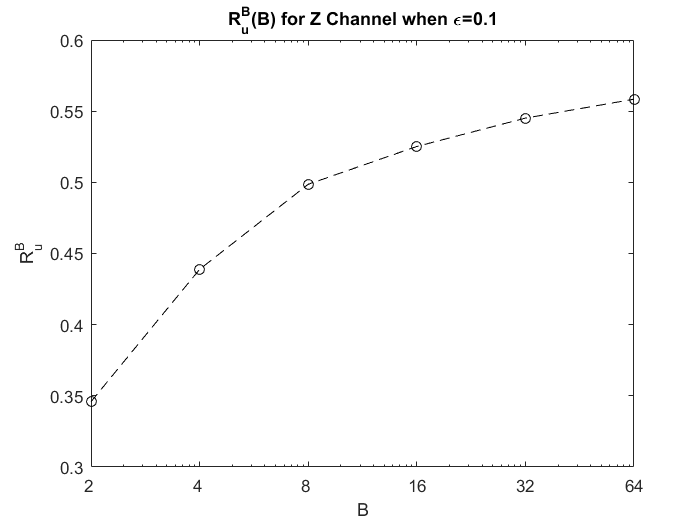}
  \captionof{figure}{\(B\) vs \(R_{u}\) in Z Channel (\(\epsilon=0.1\))}
  \label{fig:pot2}
\end{minipage}
\end{figure}

From the plots, it can be said that as \(B\) increases, \(R_{u}^B(B)\) increases. We know that, \(R_{u}^\infty\) can be expressed as a function of \(\epsilon\) and the type of channel. For BSC, \( R_{u}^\infty=(\pi ln(2))^{-1}(1-2\epsilon)^2\) and for \(\epsilon=0.01, 0.05, 0.1\), \(R_{u}^\infty=(0.441, 0.372, 0.294)\) respectively. For Z Channel \(R_{u}^\infty=(\pi ln(2)(1+\epsilon))^{-1}(1-\epsilon)\) and for \(\epsilon=0.01, 0.05, 0.1\), \(R_{u}^\infty=(0.45, 0.415, 0.375)\) respectively. If we look at the \(R_{u}^B(B)\) values for BSC and Z Channel in Figures 19, 20, 21, 22, 23, 24 we can see that \(R_{u}^B(B)\) becomes greater than \(R_{u}^\infty\) for \(B= 4, 8, 16, 32, 64\). The reasoning for that is unknown to us as we thought \(R_{u}^B(B)\) as an increasing function with respect to \(B\), when we looked at our figures. However, that might not be the case and they might be a limit where \(R_{u}^B(B)\) becomes a decreasing function and converges at \(R_{u}^\infty\).

We can compare the performances of BSC and Z Channel for GAMP decoder by looking at the \(R_{u}\) values. We know that for the same \(\epsilon\) value Z channel has higher convergence rates, because we only flip -1 values in Z channel while flip -1 and 1 values for BSC, which leads to higher SER in BSC for same \(\epsilon\) and \(R\). However, we can compare the performance of channels when their probability of error is same. For \(\epsilon=0.05\) in BSC and \(\epsilon=0.1\) in Z channel, the probability of error is 0.05 and we can see the convergence rates at Figures 21 and 24 respectively. From the figures, we can see that for the same probability of error, \(R_{u}^B(B)\) is higher for Z Channel than BSC for all B values. This is expected because our system performs better at asymmetric channels than the symmetric channels, because we can correct the errors in a superior way if the channel is asymmetric. If we compare \(R_{u}^\infty\) for those two cases, when \(\epsilon=0.05\) for BSC \(R_{u}^\infty=0.372\) while \(\epsilon=0.1\) for Z Channel \(R_{u}^\infty=0.375\) which proves the point that our decoder performs better at asymmetric channels than symmetric channels for the same probability of error.

\section{Conclusion and Future Work}

In this research, we presented an empirical study using GAMP algorithm to estimate the sparse superposition signal that is communicated through different channels at various parameter settings. The tested channels are AWGN, BSC and Z with the the changing parameters as section length \(B\), communication rate \(R\) and flip probability \(\epsilon\) or signal-to-noise ratio \(snr\). As a result, the implemented decoder works well as this can be proven by the behavior of the state evolution function.

Also, we can say that the figures of BSC and Z Channels are plotted with number of sections \(L=1000\) since bigger \(Ls\) lead to excessive computation time in Matlab using our laptops. Figures of AWGN Channels are plotted with \(L=4000\) since the computations are easier for AWGN. In all cases, the number of samples are \(10\) and Monte Carlo integrations are done with the \(MC size=10^{5}\). Only in \(B\) vs \(R_{u}\) of BSC and Z Channel, figures are plotted using the super-computer due to the very high values of \(B\) like \(32\) and \(64\).

In our project, we cannot achieve the Shannon capacity for the channels, therefore this study can be seen as sub-optimal. Spatial coupling, in conjunction with message-passing decoding, allows us to achieve Shannon capacity on memoryless channels (Donoho et al. 2013). Therefore, spatial coupling scheme can be integrated with our algorithm as a future work.

\section{Acknowledgement}
We could be able to complete this research with the helps and supervision of Jean Barbier and Mohamad Dia.

\newpage
\section{References}

\begin{enumerate}
\item Barbier, Jean. "Statistical physics and approximate message-passing algorithms for sparse linear estimation problems in signal processing and coding theory." arXiv preprint arXiv:1511.01650 (2015).
\item Rangan, Sundeep. "Generalized approximate message passing for estimation with random linear mixing." Information Theory Proceedings (ISIT), 2011 IEEE International Symposium on. IEEE, 2011.
\item Barbier, Jean, Mohamad Dia, and Nicolas Macris. "Proof of Threshold Saturation for Spatially Coupled Sparse Superposition Codes." arXiv preprint arXiv:1603.01817 (2016). 
\item Barbier, Jean, Mohamad Dia, and Nicolas Macris. "Threshold Saturation of Spatially Coupled Sparse Superposition Codes for All Memoryless Channels." arXiv preprint arXiv:1603.04591
\item Barbier, Jean, and Florent Krzakala. "Approximate message-passing decoder and capacity-achieving sparse superposition codes." arXiv preprint arXiv:1503.08040 (2015).
\item Barbier, Jean, Christophe Schülke, and Florent Krzakala. "Approximate message-passing with spatially coupled structured operators, with applications to compressed sensing and sparse superposition codes." Journal of Statistical Mechanics: Theory and Experiment 2015.5 (2015): P05013.
\item Barbier, Jean, and Florent Krzakala. "Replica analysis and approximate message passing decoder for superposition codes." Information Theory (ISIT), 2014 IEEE International Symposium on. IEEE, 2014.
\item Donoho, David L., Adel Javanmard, and Alessandro Montanari. "Information-theoretically optimal compressed sensing via spatial coupling and approximate message passing." Information Theory, IEEE Transactions on 59.11 (2013): 7434-7464.
\end{enumerate}

\end{document}